\documentclass[english,aps,prl, twocolumn,superscriptaddress]{revtex4-1}
\usepackage[utf8]{inputenc}
\usepackage{amsmath}
\usepackage{amssymb}
\usepackage{graphicx}
\usepackage{hyperref}
\usepackage{braket}
\usepackage{babel}
\usepackage{mathtools}
\usepackage{xspace}
\usepackage{xcolor}
\usepackage{physics}

\begin{document}
\title{Quantum magnetism and topological superconductivity in Yu-Shiba-Rusinov chains}

\author{Jacob F.\ Steiner}
\affiliation{\mbox{Dahlem Center for Complex Quantum Systems and Fachbereich Physik, Freie Universit\"at Berlin, 14195 Berlin, Germany}}

\author{Christophe Mora}
\affiliation{\mbox{Dahlem Center for Complex Quantum Systems and Fachbereich Physik, Freie Universit\"at Berlin, 14195 Berlin, Germany}}
\affiliation{Universit\'e  de  Paris, Laboratoire  Mat\'eriaux  et  Ph\'enom\`enes  Quantiques, CNRS, 75013  Paris,  France}

\author{Katharina J.\ Franke}
\affiliation{\mbox{Fachbereich Physik, Freie Universit\"at Berlin, 14195 Berlin, Germany}}

\author{Felix von Oppen}
\affiliation{\mbox{Dahlem Center for Complex Quantum Systems and Fachbereich Physik, Freie Universit\"at Berlin, 14195 Berlin, Germany}}

%
%\date{\today}
%
\begin{abstract}
Chains of magnetic adatoms on superconductors have been discussed as promising systems for realizing Majorana end states. Here, we show that dilute Yu-Shiba-Rusinov (YSR) chains are also a versatile platform for quantum magnetism and correlated electron dynamics, with widely adjustable spin values and couplings. Focusing on subgap excitations, we derive an extended $t-J$ model for dilute quantum YSR chains and use it to study the phase diagram as well as tunneling spectra. We  explore the implications of quantum magnetism for the formation of a topological superconducting phase, contrasting it to existing models assuming classical spin textures.
\end{abstract}
%
%\pacs{%
      %73.20.-r 	%Electron states at surfaces and interfaces
			%74.25.Jb, % Superconcuctivity: Electronic structure (photoemission, etc.)
			%74.55.+v % SC: Tunneling phenomena: single particle tunneling and STM
			%}  insert suggested PACS numbers in braces on next line
%
\maketitle 
%\maketitle must follow title, authors, abstract and \pacs

%%############################ INTRODUCTION ###############################

{\em Introduction.---}Chains of magnetic adatoms constitute a much studied platform for realizing topological superconductivity \cite{Choi2019,Pawlak2019,Flensberg2021,Jack2021}. The observation of zero-energy end states in dense Fe chains on superconducting substrates has been interpreted in terms of Majorana bound states \cite{NadjPerge2014, Ruby2015chains, Pawlak2016, Feldman2017, Jeon2017,Kim2019}. Crucial ingredients of this interpretation are the ferromagnetic order and direct hybridization between the Fe $d$ orbitals \cite{NadjPerge2014,Li2014a,Peng2015}, resulting in the formation of spin-polarized one-dimensional bands. Topological superconductivity results, when an odd number of the $d$-orbital bands cross the Fermi energy of spin-orbit coupled substrate superconductors \cite{NadjPerge2014,Li2014a,Ruby2017}. 

Dilute chains of magnetic adatoms on superconductors have been proposed as an alternative setting for topological superconductivity \cite{NadjPerge2013,Pientka2013}. In these chains, the adatoms are spaced sufficiently far that direct overlap of their $d$ orbitals is negligible, yet close enough that their Yu-Shiba-Rusinov (YSR) states \cite{Yu1965,Shiba1968,Rusinov1969,Balatsky2006} hybridize. The Ruderman-Kittel-Kasuya-Yosida (RKKY) interaction, possibly aided by magnetic anisotropy, is assumed to induce ordered magnetic textures, whose interaction with the substrate electrons has been described within a classical-spin model \cite{NadjPerge2013,Pientka2013}. In this setting, 
the YSR states of neighboring impurity spins hybridize into subgap bands, exhibiting a $p$-wave gap and topological superconductivity under suitable conditions. 

Contrasting with classical-spin models \cite{NadjPerge2013,Pientka2013,Poyhonen2014,Heimes2014,Kim2014,Brydon2015,Hoffman2015,Schecter2016,Kaladzhyan2016,Korber2018}, observations of Kondo resonances and discrete spin excitations, both on normal-metal and superconducting substrates \cite{Madhavan1998,Li1998,Hirjibehedin2007,Franke2011}, imply that individual impurity spins behave quantum mechanically. Here, we describe dilute YSR chains in terms of coupled quantum spins, show that they constitute a rich and versatile platform for quantum magnetism, and discuss its interplay with topological superconductivity. Unlike spin chains on normal-metal substrates \cite{Choi2019}, dilute YSR chains exhibit intimate coupling between spin and subgap fermionic degrees of freedom. For spin-$\frac{1}{2}$ impurities, we demonstrate that this correlated electron dynamics is described by an extension of the $t-J$ model \cite{Lee2006}, admitting topological superconductivity for ferromagnetic and spin-charge separation for antiferromagnetic RKKY coupling. Moreover, as a consequence of quantum phase transitions \cite{Sakurai1970,Oppen2021}, the effective impurity spin can deviate from the bare adatom spin due to screening by bound quasiparticles, and is in principle tunable \cite{Franke2011,Lee2014,Farinacci2018}. The RKKY coupling between impurity spins can be adjusted in strength as well as sign through the impurity spacing. Magnetic anisotropy, Dzyaloshinskii-Moriya (DM) interactions, and spin-orbit coupled substrate electrons further enrich the physics of these quantum spin chains in real materials. 

To describe the correlated electron dynamics, we project out the quasiparticle continuum of the superconductor in the limit of a large pairing gap and retain only the subgap YSR excitations induced by the magnetic adatoms. The resulting model includes a single superconducting site per adatom (and conduction-electron channel), so that Kondo renormalizations must be accounted for separately. Despite its simplicity, the model qualitatively reproduces \cite{Oppen2021} phase diagrams and excitation spectra of individual higher-spin impurities subject to single-ion anisotropy \cite{Zitko2011} and spin-$\frac{1}{2}$ dimers \cite{Zitko2010,Yao2014} obtained by the numerical renormalization group.

{\em  Model.---}Extensions of this model should thus provide a useful and tractable description of chains of spin-$S$ impurities. For a single YSR excitation per adatom, the model takes the form
\begin{eqnarray}
   &&H = \sum_j \left\{\Delta (c^\dagger_{j\uparrow}      
       c^\dagger_{j\downarrow}  +   \mathrm{h.c.} ) +  c_{j\sigma}^\dagger  [V \delta_{\sigma\sigma'} +  \mathbf{S}_{j}\! \cdot \! K
      \! \cdot \!  \mathbf{s}^{\phantom{\dagger}}_{\sigma\sigma'}] c^{\phantom{\dagger}}_{j\sigma'}  \right.
   \cr
   &&\left.   \,\,\,\,\,\,\,\,\,\,\,\,\,\,
   -t  [c^{\dagger}_{j\sigma}c^{\phantom{\dagger}}_{j+1,\sigma} + \mathrm{h.c.}] +  \mathbf{S}_j \! \cdot \! J \! \cdot \! \mathbf{S}_{j+1} +DS_{jz}^2\right\},\,\,\,\,
\label{eq:model}
\end{eqnarray}
where $j$ enumerates the adatoms along the chain and sums over repeated spin indices $\sigma,\sigma^\prime$ are implied. At each site $j$, the conduction electrons (creation operator $c^\dagger_{j\sigma}$, spin-$\frac{1}{2}$ matrices $\mathbf{s}$) couple to the 
local impurity spin $\mathbf{S}_j$ via antiferromagnetic exchange $K$ and potential scattering $V$. The hopping $t$ between adjacent superconducting sites (pairing strength $\Delta$) models the hybridization of YSR states. The quasiparticle continuum of the substrate mediates an interaction $J$ between nearest-neighbor adatom spins, incorporating both the RKKY (symmetric part) and the DM interaction (antisymmetric part) \cite{Yao2014}. (Longer-range couplings could be readily included.) Classical-spin models of YSR excitations are adapted to large $S$, subject to easy-axis anisotropy $D<0$ \cite{Oppen2021}. In general, magnetic adatoms induce YSR states in multiple conduction-electron channels. This could be included into Eq.\ (\ref{eq:model}) by coupling each impurity spin to multiple superconducting sites \cite{Oppen2021}. 

{\em Spin-$\frac{1}{2}$ impurities.---}We exemplify the physics of dilute quantum YSR chains by spin-$\frac{1}{2}$ adatoms. The individual impurities undergo a quantum phase transition with increasing exchange coupling $K$ \cite{Sakurai1970,Oppen2021,Balatsky2006}. At weak coupling, the conduction electrons are fully paired (even fermion parity), leaving an unscreened (free) impurity spin and a doubly-degenerate ground state. Within the single-site model, the doublet ground state (energy $E_\mathrm{BCS}$) takes the form $\ket{\pm}=|S^z=\ \Uparrow\! / \!\Downarrow\rangle \otimes |\mathrm{BCS}\rangle$, where $|\mathrm{BCS}\rangle = (u + v c_\downarrow^\dagger c_\uparrow^\dagger) |\mathrm{vac}\rangle$ is the paired BCS state. At strong coupling, the impurity forms a singlet with the conduction electrons by binding a quasiparticle (odd fermion parity). The resulting screened-spin ground state (energy $E_0$) is nondegenerate and takes the form $|0\rangle = \frac{1}{\sqrt{2}} (\ket{\Uparrow\downarrow}-\ket{\Downarrow\uparrow})$ \footnote{Notice that unlike for classical spins, the subgap quasiparticle states of quantum spins are not spin polarized.}. The transition between these ground states occurs when the energy of the YSR excitation $E_\mathrm{YSR} = E_0-E_\mathrm{BCS}$ changes sign (see \cite{supp} for details including explicit expressions for $u$, $v$, and $E_\mathrm{YSR}$). 

When coupling adatoms in the paired state ($E_\mathrm{YSR}>0$) into a dilute chain, they form a spin-$1/2$ chain subject to RKKY interactions. The YSR excitation of one of the adatoms into the screened state, e.g., by tunneling from an STM tip, quenches its spin and breaks its RKKY bonds. The quenched spin is mobile along the chain due to the hybridization of YSR states and propagates in a correlated spin background. In contrast, there is only a single low-energy state when coupling impurities in the screened state ($E_\mathrm{YSR}<0$). The YSR excitation of an adatom into the free-spin state introduces one free spin, which propagates in a (largely) spin-free background. These considerations assume that $|E_\mathrm{YSR}|$ is sufficiently large that the impurity spins are either all free (no bound quasiparticles) or all screened ($N$ bound quasiparticles). For intermediate values of $E_\mathrm{YSR}$, the average number of free spins varies continuously between $0$ and $N$. In fact, the Hamiltonian in Eq.\ (\ref{eq:model}) conserves only fermion parity and the total spin (for isotropic spin interactions), but not the number of conduction electrons. 

\begin{figure*}[!htb]
\centering
\includegraphics[width=0.94\textwidth]{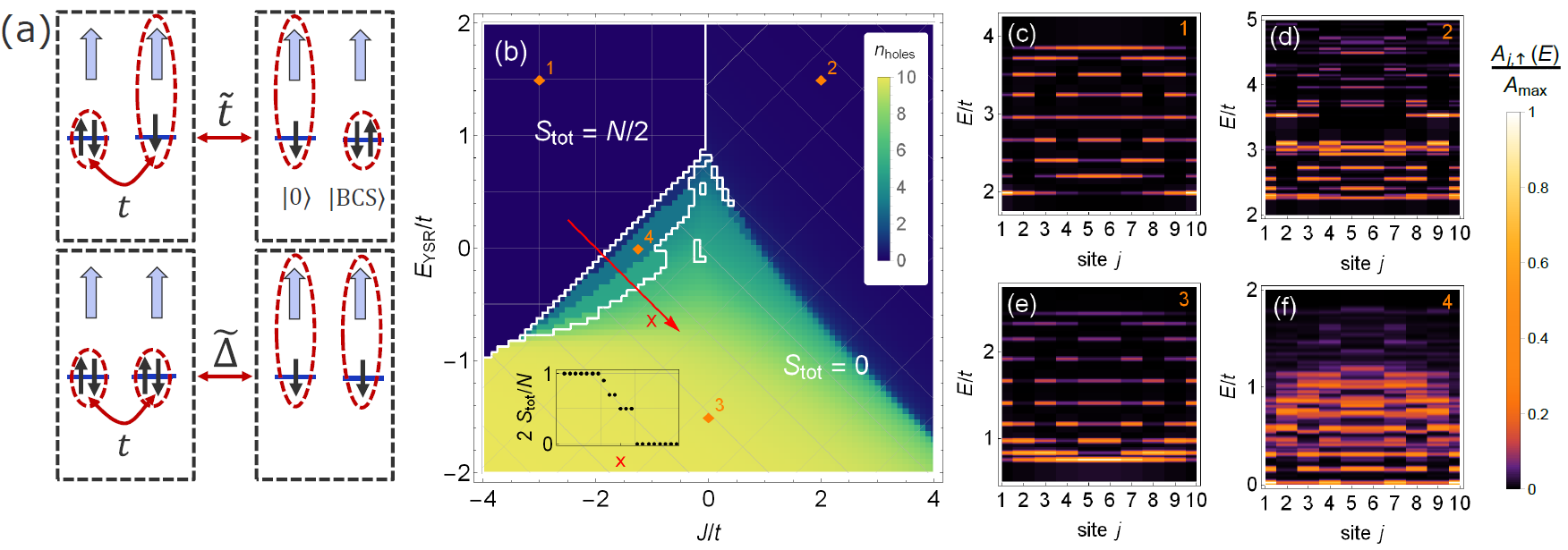}
     \caption{(a) Processes underlying effective hopping ($\tilde t$) and pairing ($\tilde\Delta$) in extended $t-J$ model. (b-f) Exact-diagonalization results for a chain of $N=10$ impurity spins ($V = 2\Delta$; $\tilde t = 2\tilde \Delta$). (b) Number of holes (screened spins) $n_{\mathrm{holes}}$ and total spin $S_\mathrm{tot}$ of a chain with periodic boundary conditions vs.\ YSR energy $E_{\mathrm{YSR}}$ and isotropic RKKY interaction $J$. For large $E_{\mathrm{YSR}}>0$, the system realizes a spin-$\frac{1}{2}$ Heisenberg chain ($n_\mathrm{holes}=0$) with ferromagnetic ($J<0$) or antiferromagnetic ($J>0$) coupling. White lines delineate borders of maximal/minimal-spin phases (horizontal/tilted mesh). Inset: Normalized total spin along the length of the arrow. (c-f) Site-resolved single-particle spectral function of chain with open boundary conditions, with 
(c) $(E_{\textrm{YSR}} , J) = (1.5,-3)$, (d) $(1.5,2)$, (e) $(-1.5,0)$, (f) $(0,-1.25)$. Panel numbers correspond with numbered diamonds in (b).}
    \label{fig:heisenberg}
\end{figure*}

{\em Mapping to extended $t-J$ model.---}We explore the full phase diagram by exact diagonalization complemented by analytical considerations. We eliminate the unphysical above-gap states, retaining only the fully paired states $\ket{\pm}$ and the singlet $\ket{0}$ of each site, by considering the limit $\Delta,K\to\infty$ at fixed YSR energy $E_\mathrm{YSR}$. In this limit, the model in Eq.\ (\ref{eq:model}) maps to an extended $t-J$ model. Regarding the local singlets $|0\rangle$ as the vacuum and introducing a spinful fermion $d_\sigma$ for each site through $\ket{\pm}=d_\pm^\dagger \ket{0}$ and $d_\pm\ket{0} =0$, we find (for details and subtleties for periodic boundary conditions, see \cite{supp}) 
\begin{eqnarray}
   && {H}_{tJ} = \sum_{j} \left\{ - E_{\textrm{YSR}}  {n}_{j} + \mathbf{{S}}_{j}\! \cdot \! J \! \cdot\! \mathbf{{S}}_{j+1} - \tilde{t} [{d}^{\dagger}_{j,\sigma} {d}^{\phantom{\dagger}}_{j+1,\sigma} + \textrm{h.c.} ]
   \right. \cr   
  &&   \left. \,\,\,\,\,\,\,\, + \tilde{\Delta} [  {d}^{\dagger}_{j,\downarrow}{d}^{\dagger}_{j+1,\uparrow} -{d}^{\dagger}_{j,\uparrow}{d}^{\dagger}_{j+1,\downarrow}  + \textrm{h.c.} ] +U {n}_{j,\uparrow} {n}_{j,\downarrow}\right\}.
\label{eq:hubbard}    
\end{eqnarray}
The projection to the physical subspace which excludes doubly-occupied sites is implemented by the limit $U\to\infty$. Here, $n_j = \sum_\sigma n_{j,\sigma} = \sum_\sigma d_{j,\sigma}^\dagger  d_{j,\sigma}^{\phantom{\dagger}}$, the spin operators $\mathbf{S}_j=\sum_{\sigma\sigma^\prime}d_{j,\sigma}^\dagger \mathbf{s}^{\phantom{\dagger}}_{\sigma\sigma'} d_{j,\sigma^\prime}^{\phantom{\dagger}}$ are now associated with the $d$ fermions, and the effective hopping and pairing amplitudes are $\tilde{t} = \frac{t}{2}(u^2-v^2)$ and $\tilde{\Delta} = tuv$. The latter originate from conduction electrons hopping between adjacent sites (amplitude $t$) as illustrated in Fig.\ \ref{fig:heisenberg}(a). Hopping between a screened and a free-spin site effectively moves the free spin (amplitude $\tilde t$) and thus the $d$ fermion. Hopping between two free-spin sites screens them (or vice versa), annihilating (creating) a $d$-fermion pair and inducing pairing (amplitude $\tilde\Delta$). Note that in classical-spin models, these amplitudes would depend sensitively on the magnetic ordering  \cite{Pientka2013,Poyhonen2014,Hoffman2015}. Finally, $E_\mathrm{YSR}$ acts as a chemical potential for the $d$-fermions.
 
The phase diagram as a function of $E_\mathrm{YSR}$ and the (isotropic) RKKY interaction $J$ can be inferred from the expectation value of the number of holes (sites with screened impurity spin) $n_{\mathrm{holes}} = N - \sum_{j} {n}_{j}$ as well as the total spin $S_\mathrm{tot}$ in the ground state. Figure \ref{fig:heisenberg}(b) shows corresponding exact diagonalization results for a chain of $N=10$ impurity spins with periodic boundary conditions. The chain has no holes for sufficiently large $E_\mathrm{YSR}>0$, realizing a spin-$\frac{1}{2}$ Heisenberg chain. As $E_\mathrm{YSR}$ is reduced, the number of holes increases and eventually becomes equal to $N$. In this state, all impurity spins are screened and the ground state has $S_\mathrm{tot}=0$. 

It is useful to contrast the transition region with the case of classical-spin textures. In the latter case, the YSR bands emerging from the positive- and negative-energy YSR states (band energies $\pm E_\mathrm{YSR}-\frac{\Lambda}{2}<\epsilon<\pm E_\mathrm{YSR}+\frac{\Lambda}{2}$) overlap for $-\frac{\Lambda}{2}<E_\mathrm{YSR}<\frac{\Lambda}{2}$. As $E_\mathrm{YSR}$ decreases from $\frac{\Lambda}{2}$ to $-\frac{\Lambda}{2}$, the number of holes $n_{\mathrm{holes}}$ increases continuously from $0$ to $N$. Here, the bandwidth $\Lambda$ is maximal for ferromagnetic textures and vanishes for antiferromagnetic ordering (for zero spin-orbit coupling). 

The behavior of the quantum chain is reminiscent of this picture for $J=0$ \footnote{The small isolated regions near $J=0$ can be understood in terms of infinite-$U$ Hubbard physics \cite{Doucot1989}.}, where $n_{\mathrm{holes}}$ varies from $0$ to $N$ for $|E_\mathrm{YSR}|\lesssim 2\tilde t$ [green region in Fig.\ \ref{fig:heisenberg}(b)]. However, we observe that increasing RKKY interaction $|J|$ stabilizes the Heisenberg chain ($n_{\mathrm{holes}}=0$) to lower $E_\mathrm{YSR}$. This shift reflects that the RKKY coupling only lowers the energy of unscreened impurity spins. On the antiferromagnetic side ($J>0$), the width of the shifted transition region [green in Fig.\ \ref{fig:heisenberg}(b)] narrows, saturating for larger $J$. While increasing antiferromagnetic correlations suppress the effect of hopping $\tilde t$, the spin-singlet pairing $\tilde \Delta$ introduces an uncertainty in $n_{\mathrm{holes}}$ which eventually governs the width of the transition region. At the same time, the total spin of the ground state is $S_\mathrm{tot}=0$, regardless of $E_\mathrm{YSR}$. As $E_\mathrm{YSR}$ is reduced, the ground state changes from a total singlet formed by antiferromagnetically coupled impurity spins to a chain of local singlets between impurity spins and conduction electrons. 

On the ferromagnetic side ($J<0$), the transition region rapidly narrows as $|J|$ increases, ultimately giving way to a direct phase boundary between Heisenberg spin chain and fully screened state. Unlike on the antiferromagnetic side, this transition is largely insensitive to the spin-singlet pairing $\tilde\Delta$ due to the strong spin polarization. At smaller $|J|$, the transition region deviates from the classical scenario, even beyond its narrowing. Reducing $E_\mathrm{YSR}$ eventually introduces holes into the Heisenberg ferromagnet, and the system becomes a metallic ferromagnet. A stepwise increase in the number of holes prompts a corresponding reduction of the total spin $S_\mathrm{tot}$ from its maximum of $N/2$ [inset in Fig.\ \ref{fig:heisenberg}(b)]. While this is similar to the classical scenario, $S_\mathrm{tot}$ jumps discontinuously to zero before the number of holes reaches $N$. Here, the metallic ferromagnet becomes energetically less favorable than a superconducting phase favored by the spin-singlet pairing $\tilde \Delta$. This singlet superconductor has $S_\mathrm{tot}=0$, as do the Heisenberg antiferromagnet and the local-singlet phase at large and negative $E_\mathrm{YSR}$. 

Pronounced differences from the classical scenario appear in the site-resolved single-particle spectral function, which can be probed directly by scanning tunneling microscopy \footnote{We compute the physical spectral function of the Hamiltonian in Eq.\ (\ref{eq:model}), which incorporates anomalous correlators of the extended $t-J$ model, see \cite{supp} for details.}. Corresponding exact-diagonalization data for a chain of $N=10$ impurity spins with open boundary conditions are shown in Fig.\ \ref{fig:heisenberg}(c-f). Probing the Heisenberg spin chain (sufficiently large $E_\mathrm{YSR}>0$), the YSR excitation of one of the impurities screens its spin and induces a mobile hole. In the ferromagnetic phase [Fig.\ \ref{fig:heisenberg}(c)], the excited hole perturbs the spin background only weakly and to a good approximation, its motion is  a single-particle problem. The local spectral function is readily understood in terms of the tight-binding spectrum of the mobile hole, when accounting for lower site energies on the boundary sites, where the hole breaks only one rather than two ferromagnetic bonds. Indeed, in the lower (upper) half of the hole band, i.e., $2\lesssim E/t \lesssim 3$ ($3\lesssim E/t \lesssim 4$), the number of nodes in the spectral function increases (decreases) with energy and there is enhanced (reduced) intensity at the ends. In the antiferromagnetic phase [Fig.\ \ref{fig:heisenberg}(d)], in contrast, the spectral weight is distributed over a much larger number of many-body states associated with the expected spin-charge separation of the antiferromagnetic $t-J$ model. 

For a chain of fully screened impurity spins [Fig.\ \ref{fig:heisenberg}(e)], the YSR excitation unscreens one of the impurity spins. To lowest order, the spectral function can again be understood in terms of a tight-binding band describing the mobile spin, now with uniform site energies throughout the chain. However, unlike in the ferromagnetic phase, the number of spins is no longer a good quantum number due to the effective pairing. The associated redistribution of spectral weight to states with additional spins leads to a reduction in intensity of the single-particle-like spectral peaks with increasing excitation energy. 

The metallic ferromagnet has strong similarities with the regime of overlapping YSR bands for classical ferromagnetic textures in the absence of spin-orbit coupling. In particular, it has a gapless excitation spectrum [Fig.\ \ref{fig:heisenberg}(f)]. The spectral function exhibits several nodes even at the lowest energy as holes are already present in the ground state, and becomes dense at higher energies due to the coupling to the particle-hole continuum. 

\begin{figure}
\centering
\includegraphics[width=.95\columnwidth]{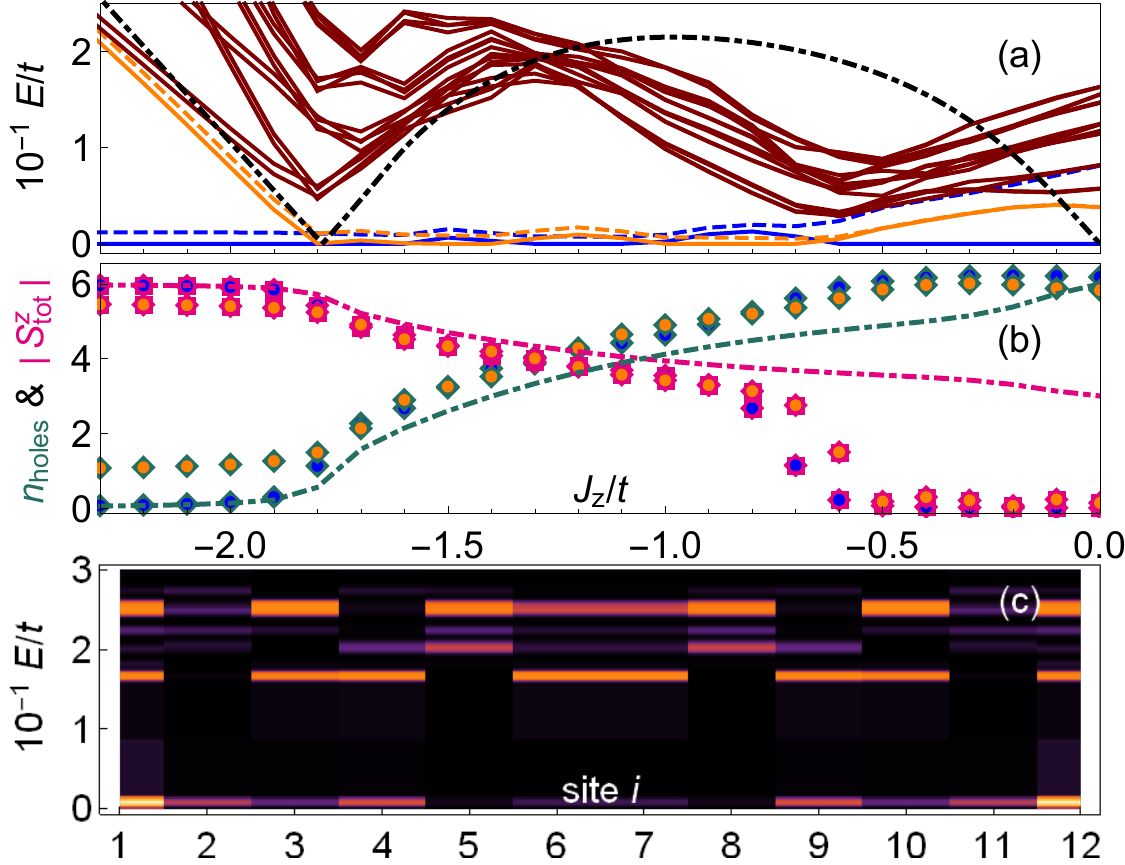}
     \caption{Exact-diagonalization results for $N=12$ chain with Ising RKKY interaction and spin-orbit coupling. (a) Ten lowest-energy eigenstates each vs.\ RKKY coupling $J_z$
in even- and odd-parity sectors at $E_\mathrm{YSR}=0$, cutting through the metallic ferromagnet for zero spin-orbit coupling (for phase diagram, see \cite{supp}). 
For $-1.8\lesssim J_z/t\lesssim -0.7$, the four lowest states (even/odd parity: blue/orange) are separated by a topological gap from a continuum of excited states. 
Black dash-dotted line: Approximate analytical gap \cite{supp}. (b) Corresponding $n_{\textrm{holes}}$ (green, diamonds) and total magnetization $\abs{ S^z_\mathrm{tot} }$ (pink, stars). Dash-dotted lines: Analytical results based on trial state. Fillings of symbols color coded as in (a). (c) Tunneling spectra at $J_z=-1.25t$ revealing zero-energy Majorana end states. (Parameters: $V = 2\Delta$, spin-orbit coupling $\alpha = 0.25$; a minute Zeeman field $B_z = 10^{-3}t$ singles out spin-polarized states in numerics and induces small intra-parity splittings in ground-state manifold.)}
    \label{fig:ising_with_soc}
\end{figure}

The metallic ferromagnet can become a topological superconductor for spin-orbit coupled superconductors (as for classical spin textures). Spin-orbit coupling makes the YSR hybridization spin dependent and breaks spin-rotation symmetry.
Specifically, we introduce spin-dependent hopping $-t\sum_j c_{j}^\dagger (1+ i\alpha\sigma_y) c^{\phantom{\dagger}}_{j+1} + \mathrm{h.c.}$ and anisotropic (Ising-like) RKKY interactions $J\sum_j S_j^zS_{j+1}^z$ polarized perpendicular to the spin-orbit field. Without double-occupation constraint, this model was studied in Ref.\ \cite{Stoudenmire2011} as a paradigm for the interplay of topological superconductivity and interactions. 

Corresponding numerical results are shown in Fig.\ \ref{fig:ising_with_soc} for a chain of $N=12$ sites with open boundary conditions. Without spin-orbit coupling, the phase diagram (see \cite{supp}) is qualitatively similar to Fig.\ \ref{fig:heisenberg}(b) for Heisenberg interactions. Spin-orbit coupling only weakly affects the ferromagnetic insulator at 
$J_z/t\lesssim -1.8$ or the singlet superconductor at $J_z/t\gtrsim -0.7$, but the spectrum of the metallic ferromagnet ($-1.8\lesssim J_z/t\lesssim -0.7$) develops a $p$-wave pairing gap [Fig.\ \ref{fig:ising_with_soc}(a)]. The associated formation of Majorana end states leads to four (up to finite-size corrections) degenerate ground states, a pair of even- and odd- fermion-parity states for each of the two symmetry-broken spin configurations. 
In line with a topological degeneracy, even- and odd-parity ground states are indistinguishable by the local observables $n_{\textrm{holes}}$ and the total spin projection $S^z_\mathrm{tot}$ [Fig.\ \ref{fig:ising_with_soc}(b)]. For sufficiently large $|J|$, the $p$-wave gap [Fig.\ \ref{fig:ising_with_soc}(a)] as well as $n_\mathrm{holes}$ and $S^z_\mathrm{tot}$ [Fig.\ \ref{fig:ising_with_soc}(b)] can be well reproduced analytically, using a variational trial state for the metallic ferromagnet and including the spin-orbit coupling into the extended $t-J$ model (see \cite{supp} for theoretical details). At smaller $|J|$, the analytical description breaks down as it neglects the effects of the singlet pairing $\tilde \Delta$. Tunneling spectra clearly reveal the formation of zero-energy Majorana end states protected by a gap [Fig.\ \ref{fig:ising_with_soc}(c)].  

{\em Discussion.---}Dilute YSR chains constitute a versatile platform for quantum magnetism. Even for spin-$\frac{1}{2}$ adatoms, we uncover a rich phase diagram described by an extended $t-J$ model. Unlike the standard $t-J$ model, there are no restrictions on the sign of $J$ nor on its strength relative to $t$. Tunneling spectra reflect its local spectral function, when accounting for additional pairing correlations. 

Spin-$\frac{1}{2}$ impurities are directly realized for magnetic adatoms with one unpaired electron in the valence shell (e.g., Cerium). Importantly, however, the relevant spin is not identical to the bare spin $S_0$ of the magnetic adatom on the superconducting substrate. Adatoms can bind a quasiparticle in any of the $2S_0$ conduction-electron channels. The effective spin is thus equal to $S_0-Q/2$, where $Q$ denotes the number of bound quasiparticles, and the extended $t-J$ model can apply to higher-spin impurities, if all but one channel robustly bind a quasiparticle. 

More generally, the extended $t-J$ model exemplifies the importance of treating dilute YSR chains as quantum spin chains. Their effective spin depends on the detailed coupling between adatom and substrate, and can conceivably be tuned even for a given system, for instance on gate-tunable superconductors \cite{Lee2014} and on moir\'e \cite{Franke2011} or charge-density-modulated \cite{Liebhaber2019} structures. The phenomenology of dilute YSR chains with higher spins is further refined by single-ion anisotropy as well as both intra- and interchannel YSR hybridization. 

Unraveling the phenomenology of quantum YSR chains therefore promises important insights into the physics of magnetic adatoms on superconductors. While hybridizing subgap states form YSR bands in models with classical spin textures, we find that subgap spectra exhibit a plethora of qualitatively distinct behaviors depending on the magnetic phase. The quantum magnetism must also inform the search for topological superconductivity and Majorana zero modes in dilute YSR chains. In particular, we find that the parent metallic ferromagnet is limited in scope by competing insulating-ferromagnet and singlet-superconductor phases. 

Finally, the physics of dilute YSR chains is not limited to magnetic adatoms, but can also be realized in chains of Coulomb-blockaded quantum dots coupled to a superconductor. Previous theoretical work focused on classical spins \cite{Choi2011,Sau2012,Fulga2013}, but recent experiments on double quantum dots probe the elementary unit of quantum YSR chains \cite{Grove2018}. This setting naturally realizes the spin-$\frac{1}{2}$ case of the extended $t-J$ model and provides a promising complement to recent work \cite{Dehollain2020} on quantum dot arrays as quantum simulators of quantum magnetism.

\begin{acknowledgments}
We acknowledge discussions with E.\ Liebhaber, G.\ Reecht, and L.\ M.\ R\"utten, and are grateful for funding through QuantERA grant TOPOQUANT and by Deutsche Forschungsgemeinschaft through CRC 183 (project C03) as well as CRC 910. 
\end{acknowledgments}
  
%\bibliographystyle{apsrev4-1}
%\bibliography{library}


\onecolumngrid

\clearpage

\setcounter{figure}{0}
\setcounter{section}{0}
\setcounter{equation}{0}
\renewcommand{\theequation}{S\arabic{equation}}
\renewcommand{\thefigure}{S\arabic{figure}}

\onecolumngrid

\section*{\Large{Supplemental Material}}

\section{Mapping to extended $t-J$ model}\label{sec:supp_mapping_to_t-J}

In this section, we present details of the mapping from the original model in Eq.\ \eqref{eq:model} to the extended $t-J$ model in Eq.\ \eqref{eq:hubbard}, which is valid for impurities with spin $S=1/2$. We will first discuss the mapping for a single impurity (Sec.\ \ref{sec:supp_single_impurity}) and then extend it to a chain of impurities (Sec.\ \ref{sec:supp_spin_chain}). Finally, we include spin-orbit coupling (Sec.\ \ref{sec:supp_ext_t-J_soc}) and consider the spectral function (Sec.\ \ref{sec:supp_spectral_function}) at the level of the extended $t-J$ model. 

We begin by briefly summarizing the main ideas. As described in the main text, the model in Eq.\ \eqref{eq:model} has three low-energy states for each site: the singlet state $\ket{0} = \frac{1}{\sqrt{2}}\pqty{\ket{\Uparrow\downarrow} - \ket{\Downarrow\uparrow} }$ corresponding to a screened impurity spin, and the BCS-paired doublet $\ket{\pm} = \ket{S^z = \pm \frac{1}{2}}\otimes \ket{\textrm{BCS}}$ describing a free impurity spin. This resembles a single spinful electronic orbital with infinite on-site repulsion, when considering the spin singlet $|0\rangle$ as the empty orbital and the spin-up and spin-down states $|\pm\rangle$ of the unscreened doublet as the singly-occupied orbital. We project out the doubly-occupied site, which has no analog in the original model, by including an infinite on-site potential. 

The fermionic operators $d_{\sigma}$ for this orbital are introduced as follows. The singlet involves the singly-occupied states $|\sigma\rangle = c^\dagger_\sigma|\mathrm{vac}\rangle$, which can also be obtained from the BCS ground state using the Bogoliubov operators $\gamma_\sigma$ through $|\sigma\rangle = \gamma_\sigma^\dagger |\mathrm{BCS}\rangle$. The free-spin states (involving $|\mathrm{BCS}\rangle$ as the electronic state) are thus obtained from the singlet by annihilating a quasiparticle, and the new fermion creation operators $d_\sigma^\dagger$ essentially correspond to the quasiparticle annihilation operators $\gamma_{\bar{\sigma}}$. Taking into account the detailed singlet state, we find (restoring the site index)
\begin{align}\label{eq:hubbard_mapping}
    \gamma_{j,\sigma} =  u {c}_{j,\sigma} + \sigma v {c}^{\dagger}_{j,\bar{\sigma}} \sim \sigma \frac{(-1)^{j}}{\sqrt{2}} d^{\dagger}_{j,\bar{\sigma}}.
\end{align}
Here, we use $\sim$ rather than an equality since strictly speaking, the two operators are 
equivalent only when projecting to the low-energy subspace. 

Importantly, the $d$-fermions of the extended $t-J$ model therefore correspond to Bogoliubov operators of the original Hamiltonian. This implies that the electron spectral function measured in tunneling experiments is not identical to the spectral function of the extended $t-J$ model. Instead, it involves anomalous terms when written in terms of the $d$-fermions. 

Below, we consider the mapping as well as the physical spectral function in detail. 

\subsection{Single magnetic impurity}\label{sec:supp_single_impurity}

We first consider a single $S=1/2$ impurity coupled to a single-site superconductor as described by the Hamiltonian 
\begin{equation}
  H = \Delta \pqty{c^\dagger_{\uparrow} c^\dagger_{\downarrow} + \mathrm{h.c.}} +\sum_{\sigma\sigma'} c_{\sigma}^\dagger  \pqty{ V \delta_{\sigma\sigma'} +  K \mathbf{S}\cdot
      \mathbf{s}_{\sigma\sigma'}} c_{\sigma'}
\end{equation}
and focus on YSR states in the vicinity of the band center. We can formally implement this limit by taking the exchange coupling $K$ (assumed isotropic), the gap $\Delta$, and the potential scattering $V$ of the substrate to infinity, while keeping the energy of the Yu-Shiba-Rusinov (YSR) state and the particle-hole asymmetry $V/\Delta$ fixed. 

The Hamiltonian $H$ conserves fermion parity. Without coupling to the impurity spin, the even-fermion-parity eigenstates and eigenenergies of the single-site superconductor are
\begin{subequations}\label{eq:energies_and_states_bcs_even}
\begin{align}
    \ket{\textrm{BCS}} =&\  u  \ket{\textrm{vac}} +  v \ket{\downarrow\uparrow} ,\ E_{\textrm{BCS}} = V - \sqrt{\Delta^2 + V^2},  \label{eq:BCS-state} \\
    \ket{\overline{\textrm{BCS}}} =&\ v  \ket{\textrm{vac}} -  u \ket{\downarrow\uparrow},\ E_{\overline{\textrm{BCS}}} = V + \sqrt{\Delta^2 + V^2}.
\end{align}
\end{subequations}
Here, $\ket{\downarrow\uparrow} =  {c}^{\dagger}_{\downarrow}  {c}^{ \dagger}_{\uparrow}  \ket{\textrm{vac}}$. (We reserve $\ket{0}$ for the singlet state.) $\ket{\textrm{BCS}}$ denotes the BCS ground state and $\ket{\overline{\textrm{BCS}}}$ the excited state with two Boguliubov quasiparticles. The odd-fermion-parity eigenstates 
\begin{align}\label{eq:energies_and_states_bcs_odd}
    \ket{\sigma} = {c}^{\dagger}_{\sigma }  \ket{\textrm{vac}} = {\gamma}^{\dagger}_{\sigma} \ket{\textrm{BCS}},\ E_{\sigma} = V
\end{align}
can be either viewed as excited states with one Bogoliubov quasiparticle or as single-electron states unaffected by pairing. Here, we defined the Bogoliubov quasiparticle operators 
\begin{align}\label{eq:bogoliubov_operators}
   {\gamma}^{\dagger}_{\uparrow} = u {c}^{\dagger}_{\uparrow} + v {c}_{\downarrow},\ 
   {\gamma}^{\dagger}_{\downarrow} = u {c}^{\dagger}_{\downarrow} - v {c}_{\uparrow},
\end{align} 
with the electron and hole amplitudes
\begin{align}\label{eq:u_and_v}
    u = \sqrt{\frac{1}{2}\pqty{1 + \frac{V}{\sqrt{\Delta^2 + V^2}}}},\ 
    v = \sqrt{\frac{1}{2}\pqty{1 - \frac{V}{\sqrt{\Delta^2 + V^2}}}}.
\end{align}
The electron and hole amplitudes are in general different as a result of the potential scattering $V$ by the impurity.

The exchange interaction $K$ couples the single-site superconductor to the impurity spin. Due to Eq.\ \eqref{eq:energies_and_states_bcs_odd}, we can replace the electron operators $c_\sigma$ in the exchange Hamiltonian by the corresponding Bogoliubov operators.  Expressing the Hamiltonian of the single-site superconductor in terms of the Bogoliubov operators, we have
\begin{equation}
    H = V + \sqrt{\Delta^2 + V^2}\pqty{\sum_{\sigma} {n}_{\sigma} - 1} +\sum_{\sigma\sigma'} K \gamma_{\sigma}^\dagger  \mathbf{S}\cdot
      \mathbf{s}_{\sigma\sigma'} \gamma_{\sigma'},
\end{equation}
where ${n}_{\sigma} = {\gamma}_{\sigma}^{\dagger} {\gamma}_{\sigma}$. We can also write the exchange coupling in terms of spin raising and lowering operators, $\mathbf{S}\cdot \mathbf{s}={S}^z {s}^z + \frac{1}{2} ({S}^+ {s}^- +  {S}^- {s}^+)$. In the following, we will at times use the abbreviated notation
\begin{align}
    {s}^z =&\ \frac{1}{2}\pqty{{c}^{\dagger}_{\uparrow} {c}_{\uparrow} - {c}^{\dagger}_{\downarrow}  {c}_{\downarrow} } = \frac{1}{2}\pqty{{n}_{\uparrow} - {n}_{\downarrow}},\ 
    {s}^+ = \pqty{s^-}^{\dagger} =  {c}^{\dagger}_{\uparrow}  {c}_{\downarrow} = {\gamma}^{\dagger}_{\uparrow}  {\gamma}_{\downarrow},
\end{align}
which incorporates the fermion operators into $\mathbf{s}$. The even-fermion-parity sector is not affected by the exchange coupling as all corresponding matrix elements of $\mathbf{s}$ vanish. As a result, the even-fermion-parity states are doubly degenerate due to the unscreened impurity spin. This can also be viewed as a Kramers degeneracy. The ground states in the even-fermion-parity sector are thus 
\begin{equation}
	\ket{\pm} \equiv \ket{S^z = \pm \frac{1}{2}, \textrm{BCS}} = \ket{S^z = \pm \frac{1}{2}} \otimes \ket{\textrm{BCS}}
\end{equation}
with energy $E_{\textrm{BCS}}$.
Within the odd-fermion-parity sector, the antiferromagnetic exchange couples the electron and impurity spins into a singlet ground state,
\begin{align}\label{eq:singlet}
    \ket{0} = \frac{1}{\sqrt{2}} ( \ket{\Uparrow \downarrow} - \ket{\Downarrow \uparrow}),\  E_{0} = V - \frac{3}{4} K.
\end{align}
where $\ket{S^z, \sigma} = \ket{S^z} \otimes {c}^{\dagger}_{\sigma}\ket{\textrm{vac}}$.  
The excited triplet states have energy $V + K/4$. Comparing the ground-state energies of the even- and odd-fermion-parity ground states in Eqs.\ (\ref{eq:BCS-state}) and (\ref{eq:singlet}), respectively, one concludes that there is a quantum phase transition between even- and odd-fermion-parity ground states when $E_{\textrm{BCS}} = E_{0}$, i.e., when 
\begin{equation}
    \frac{3}{4} K = \sqrt{\Delta^2 + V^2 }.
\end{equation}
The energy difference between these ground states is the excitation energy of the Yu-Shiba-Rusinov state,
\begin{equation}\label{eq:energy_ysr}
    E_{\textrm{YSR}} = E_0 - E_{\textrm{BCS}} = \sqrt{\Delta^2 + V^2 } - \frac{3}{4} K. 
\end{equation}
We thus define $E_{\textrm{YSR}}$ as positive when the impurity spin in unscreened in the ground state and as negative when the impurity is screened. 

We now focus on the limit of $K$, $\Delta$, and $V$ large, keeping $E_{\textrm{YSR}}$ and $V/\Delta$ fixed, thereby eliminating above-gap excitations (which are not properly described within the single-site model due to the lack of a quasiparticle continuum). Then, only the singlet state and the BCS states with unscreened impurity spin remain relevant for the low-energy physics. We can define the projector onto the low-energy subspace through
\begin{equation}\label{eq:single_site_proj}
    P = P_{\textrm{BCS}} + P_0, \textrm{ with } P_{\textrm{BCS}} = \sum_{\pm} \ketbra{\pm}{\pm},\ P_0 = \ketbra{0}{0}.
\end{equation}
The low-energy projected Hamiltonian is then
\begin{equation}
 P H P = -E_{\textrm{YSR}} P_{\textrm{BCS}} + \pqty{V - \frac{3}{4}K} P. 
\end{equation}
We proceed by investigating the action of the projected Bogoliubov operators $P \gamma_{\sigma} P$, $P \gamma^{\dagger}_{\sigma} P$ as well as $P S^i P$ on the low-energy subspace. We will find that they satisfy the same algebra as the $t-J$ fermions (i.e., spinful electrons with the constraint that double occupancy is forbidden).
Consider first the action of $P \gamma_{\sigma} P$ on the singlet, 
\begin{align}
       P \gamma_{\downarrow} P \ket{0} = \frac{1}{\sqrt{2}} \ket{+},\ 
       P \gamma_{\uparrow}  P \ket{0} =  -\frac{1}{\sqrt{2}} \ket{-}.
\end{align}
Thus, the states $\ket{\pm}$ are created from $\ket{0}$ by application of $P \gamma_{\downarrow / \uparrow} P$ (up to normalization). Similarly, we have
\begin{align}  
      P \gamma^{\dagger}_{\sigma} P \ket{\pm} =  \begin{cases} 0 \textrm{ if } \sigma = \pm  \\  \frac{\bar{\sigma}}{\sqrt{2}} \ket{0} \textrm{ if } \sigma = \mp \end{cases} ,\  
      P \gamma^{\dagger}_{\sigma} P \ket{0} = 0.
\end{align}
Thus, $P \gamma^{\dagger}_{\sigma} P$ acts as annihilator for the excitations with $\pm = \bar{\sigma}$, while $\ket{0}$ acts as the vacuum. Finally, consider the action of $P \gamma_{\sigma} P$ on the already occupied $\ket{\pm} $,
\begin{align}
      P \gamma_{\sigma}  P \ket{\pm} =&\ 0.
\end{align}
Following these observations, for $\sigma = \mp$ this corresponds to adding a second spin-$\sigma$ electron into the already occupied state, which should indeed give $0$. However, the above also implies $P \gamma_{\uparrow} P \gamma_{\downarrow} P \ket{0} = 0$, which corresponds to forbidden double occupation. 
This motivates the definition of new quasi-fermionic creation operators
\begin{align}
    \Gamma_{\sigma}^{\dagger} =&\ \sigma \sqrt{2} P \gamma_{\bar{\sigma}}  P,
\end{align}
which create the states $\ket{\pm = \sigma}$ from $\ket{0}$. They satisfy
\begin{align}
    \Gamma_{\sigma}^{\dagger}\Gamma_{\sigma}^{\dagger} = 0 =  \Gamma_{\sigma}\Gamma_{\sigma}.
\end{align}
Due to the forbidden double occupation, these are not proper fermions. Their full anticommutation algebra on the low-energy subspace is given by
\begin{subequations}
\begin{align}
    \Bqty{\Gamma_{\uparrow},\Gamma_{\uparrow}^{\dagger}} \begin{cases} \ket{0} \\ \ket{+} \\ \ket{-} \end{cases}  =&\ \begin{cases} \ket{0} \\ \ket{+} \\ 0 \end{cases},\ 
    \Bqty{\Gamma_{\downarrow},\Gamma_{\downarrow}^{\dagger}} \begin{cases} \ket{0} \\ \ket{+} \\ \ket{-} \end{cases}  = \begin{cases} \ket{0} \\ 0 \\ \ket{-} \end{cases},\\
    \Bqty{\Gamma_{\downarrow},\Gamma_{\uparrow}^{\dagger}} \begin{cases} \ket{0} \\ \ket{+} \\ \ket{-} \end{cases}  =&\ \begin{cases} 0 \\ 0 \\ \ket{+} \end{cases},\ 
    \Bqty{\Gamma_{\uparrow},\Gamma_{\downarrow}^{\dagger}} \begin{cases} \ket{0} \\ \ket{+} \\ \ket{-} \end{cases}  = \begin{cases} 0 \\  \ket{-}  \\ 0\end{cases},
\end{align}
while
\begin{align}
    \Bqty{\Gamma_{\uparrow}^{\dagger},\Gamma_{\downarrow}^{\dagger}} = 0 = \Bqty{\Gamma_{\uparrow},\Gamma_{\downarrow}}.
\end{align}
\end{subequations}
We now show explicitly that these operators satisfy the same algebra as the $t-J$ fermions. To this end, consider a single site, infinite-$U$ Hubbard model described by fermionic operators $d_{\sigma}$, with states $\ket{\textrm{vac}_{\textrm{H}}}, \ket{\sigma_{\textrm{H}}} = d^{\dagger}_{\sigma} \ket{\textrm{vac}_{\textrm{H}}}$ and the doubly occupied $\ket{\downarrow\uparrow_{\textrm{H}}} = d^{\dagger}_{\downarrow}d^{\dagger}_{\uparrow} \ket{\textrm{vac}_{\textrm{H}}}$, as well as the projector onto the empty and singly-occupied states, $P_{\textrm{H}} = 1 - \ketbra{\downarrow\uparrow_{\textrm{H}}}{\downarrow\uparrow_{\textrm{H}}}$. Note that, for sake of simplicity, in the main text no distinction is made between the Hilbert space of the original model and that of the Hubbard model. As above we define the projected quasi-fermion operators as 
\begin{equation}
    D_{\sigma} = P_{\textrm{H}} d_{\sigma} P_{\textrm{H}}.
\end{equation}
It is straightforward to check the anticommutation relations. As above
\begin{subequations}
\begin{align}
    \Bqty{D_{\uparrow},D_{\uparrow}^{\dagger}} \begin{cases} \ket{\textrm{vac}_{\textrm{H}}} \\ \ket{\uparrow_{\textrm{H}}} \\ \ket{\downarrow_{\textrm{H}}} \end{cases}  =&\ \begin{cases} \ket{\textrm{vac}_{\textrm{H}}} \\ \ket{\uparrow_{\textrm{H}}} \\ - P_{\textrm{H}}  d_{\uparrow} P_{\textrm{H}} \ket{\downarrow\uparrow_{\textrm{H}}} = 0 \end{cases}, \\ 
    \Bqty{D_{\downarrow},D_{\downarrow}^{\dagger}} \begin{cases} \ket{\textrm{vac}_{\textrm{H}}} \\ \ket{\uparrow_{\textrm{H}}} \\ \ket{\downarrow_{\textrm{H}}} \end{cases}  =&\  \begin{cases} \ket{\textrm{vac}_{\textrm{H}}} \\ P_{\textrm{H}} d_{\downarrow} P_{\textrm{H}} \ket{\downarrow\uparrow_{\textrm{H}}} = 0 \\ \ket{\downarrow_{\textrm{H}}} \end{cases},\\
    \Bqty{D_{\downarrow},D_{\uparrow}^{\dagger}} \begin{cases} \ket{\textrm{vac}_{\textrm{H}}} \\ \ket{\uparrow_{\textrm{H}}} \\ \ket{\downarrow_{\textrm{H}}} \end{cases} =&\ \begin{cases} 0 \\ 0  \\ - P_{\textrm{H}} D_{\downarrow} d_{\textrm{H}} \ket{\downarrow\uparrow_{\textrm{H}}} + \ket{\uparrow_{\textrm{H}}} = \ket{\uparrow_{\textrm{H}}} \end{cases},\\
    \Bqty{D_{\uparrow},D_{\downarrow}^{\dagger}} \begin{cases} \ket{\textrm{vac}_{\textrm{H}}} \\ \ket{\uparrow_{\textrm{H}}} \\ \ket{\downarrow_{\textrm{H}}} \end{cases} =&\  \begin{cases} 0 \\ P_{\textrm{H}} d_{\uparrow} P_{\textrm{H}} \ket{\downarrow\uparrow_{\textrm{H}}} + \ket{\downarrow_{\textrm{H}}} = \ket{\downarrow_{\textrm{H}}}  \\ 0\end{cases},
\end{align}
and
\begin{align}
    \Bqty{D_{\uparrow}^{\dagger},D_{\downarrow}^{\dagger}} = 0 = \Bqty{D_{\uparrow},D_{\downarrow}}.
\end{align}
\end{subequations}
Thus, the low-energy projected superconductor may be described using the infinite-$U$ Hubbard model instead, with $\gamma_{\sigma} \sim  -\sigma d^{\dagger}_{\bar{\sigma}}/\sqrt{2}$. As mentioned above, we introduce the $\sim$-symbol to mean equivalence at the level of the low-energy projected theory. The corresponding Hamiltonian is 
\begin{equation}
	H_{\textrm{H}} = -E_{\textrm{YSR}} \sum_{\sigma} d^{\dagger}_{\sigma} d_{\sigma} + V - \frac{3}{4}K + U  d_{\uparrow}^{\dagger} d_{\uparrow}  d_{\downarrow}^{\dagger} d_{\downarrow},\ U \to \infty .
\end{equation}
The merits of this identification will become clear when considering chains in the next section.

We have not yet discussed the low-energy projected spin operators $P S^i P$. They act like the spin operators associated with the $d$ fermions,
\begin{equation}
S^z \sim \frac{1}{2}\sum_{\sigma} \sigma d^{\dagger}_{\sigma} d_{\sigma},\  S^+  \sim d_{\uparrow}^{\dagger}d_{\downarrow},\ S^- \sim d_{\downarrow}^{\dagger}d_{\uparrow}. 
\end{equation}
To see this, we explicitly consider their action on the low-energy states,
\begin{subequations}
\begin{align}
     P S^z P \ket{0} \propto&\ P \pqty{ \ket{\Uparrow \downarrow} + \ket{\Downarrow \uparrow} } = 0, \ 
     P S^z P \ket{\pm} = \pm \frac{1}{2} \ket{\pm},  \\ 
     P S^+ P \ket{0} \propto&\ P  \ket{\Uparrow \uparrow} = 0, \ 
     P S^{\pm} P \ket{\pm} = 0,\ 
     P S^{\mp} P \ket{\pm} = \ket{\mp}.
\end{align}
\end{subequations}

\subsection{Spin-$\frac{1}{2}$ chain}\label{sec:supp_spin_chain}

We now generalize to a chain of spin-$\frac{1}{2}$ impurities with isotropic exchange $K$ and general RKKY interaction $J$ (possibly including Dzyaloshinskii-Moriya interactions),
\begin{align}
   H = \sum_j \bqty{ \Delta \pqty{ c^\dagger_{j,\uparrow}      
       c^\dagger_{j,\downarrow}  +   \mathrm{h.c.} } + \sum_{\sigma\sigma'} c_{j,\sigma}^\dagger  \pqty{ V \delta_{\sigma\sigma'} +  K \mathbf{S}_{j}
      \cdot   \mathbf{s}_{\sigma\sigma'} } c_{j,\sigma'} -t \sum_\sigma \pqty{ c^{\dagger}_{j,\sigma}c_{j+1,\sigma} + \mathrm{h.c.} }
      +  \mathbf{S}_j \cdot J \cdot \mathbf{S}_{j+1} }.
\end{align}
Note that there is no single-ion anisotropy for spin-$\frac{1}{2}$ impurities. 

We want to project this Hamiltonian to the low-energy subspace introduced in the previous section. As a first step, we rewrite the tunneling Hamiltonian in terms of Bogoliubov operators,
\begin{align}
  H_t =  \sum_{j}\bqty{ -\frac{t}{2}\pqty{u^2 - v^2} \sum_\sigma\pqty{ \gamma^{\dagger}_{j,\sigma} \gamma_{j+1,\sigma} + \textrm{h.c.} }
 + t uv \pqty{ \gamma^{\dagger}_{j,\uparrow} {\gamma}^{\dagger}_{j+1,\downarrow} - {\gamma}^{\dagger}_{j,\downarrow} {\gamma}^{\dagger}_{j+1,\uparrow}  + \textrm{h.c.}} },
\end{align}
and define the effective hopping and pairing amplitudes
\begin{equation}
	\tilde{t} = \frac{t}{2}(u^2 - v^2)  = \frac{t V}{2 \sqrt{\Delta^2 + V^2}},\ 
	\tilde{\Delta} = t u v  = \frac{t \Delta}{2 \sqrt{\Delta^2 + V^2}}. 
\end{equation}
Next, we apply the projection
\begin{equation}
    \mathcal{P} = \prod_j P_j, 
\end{equation}
where $P_j$ are the local projectors defined by Eq.\ \eqref{eq:single_site_proj}. We obtain
\begin{align}
   \mathcal{P}H\mathcal{P} =&\ \mathcal{P} \sum_j  \Bigg[ -E_{\textrm{YSR}} P_{j,\textrm{BCS}} + \pqty{V - \frac{3}{4}K} P_{j}  - \tilde{t} \sum_\sigma\pqty{ \Gamma_{j,\sigma} \Gamma^{\dagger}_{j+1,\sigma} + \textrm{h.c.} }
   \nonumber
 \\ &\ \ \ \ \ \ \ \ \ \ \ 
 + \tilde{\Delta} \pqty{ - \Gamma_{j,\downarrow} \Gamma_{j+1,\uparrow} + \Gamma_{j,\uparrow} \Gamma_{j+1,\downarrow}  + \textrm{h.c.}} 
      +  P_j \mathbf{S}_j P_j \cdot J \cdot P_{j+1}\mathbf{S}_{j+1} P_{j+1} \Bigg] \mathcal{P},
\end{align}
in terms of the projected operators $\Gamma_{j,\sigma} = \sigma \sqrt{2} P_j \gamma^{\dagger}_{j,\bar{\sigma}} P_j$. As in the single-site case discussed in the previous section, the $\Gamma_{j,\sigma}$ have the same anticommutation algebra as $t-J$ fermions. The on-site algebra follows from the previous section. Furthermore, it is clear that $\{ \Gamma_{j,\sigma} , \Gamma_{j',\sigma'} \} \propto P_j P_{j'} \{ \gamma^{\dagger}_{j,\bar{\sigma}},  \gamma^{\dagger}_{j',\bar{\sigma}'} \} P_{j}' P_{j} = 0 $ for $j \neq j'$ since $[ P_j , P_{j'} ] =0$ and similarly for the anticommutator of creation and annihilation operators. 
 
There is one more issue to account for to make the mapping exact. The new vacuum is a product of local singlet states, and therefore has odd local fermion parities. Similarly, the single-occupied sites actually have even local fermion parity. Thus, acting with $\Gamma^{\dagger}_j$ on a basis state, there is an overall minus sign equal to the parity of the number of singlets on sites $i<j$, i.e. of the number of ``unoccupied" rather than  ``occupied" states. To make this precise, consider the action of $\Gamma_{j,\sigma}$ on the product basis derived from $\ket{0}$ and $\ket{S_z, \textrm{BCS}}$. We write 
\begin{align}
    \ket{\mathbf{y}} \equiv \ket{y_1} \otimes ... \otimes \ket{y_N}
\end{align}
in terms of $y_j \in \{ -,0,+ \}$. 
Then, we have
\begin{align}
    \Gamma_{j,\sigma}\ket{\mathbf{y}} =&\ (-1)^{\sum_{i<n}(1+\sum_{\sigma'} n_{i,\sigma'})}\nonumber \\ 
    &\ \ \ \ \ \ \ \ \ \ket{y_1} \otimes...\otimes \Gamma_{j,\sigma}\ket{y_j} \otimes ...  \otimes \ket{y_N}. 
\end{align}
Similarly, for the infinite-$U$ Hubbard model
\begin{align}
    D_{j,\sigma}\ket{\mathbf{x}} =&\ (-1)^{\sum_{i<n,\sigma'} n_{i,\sigma'}}\nonumber \\ 
    &\ \ \ \ \ \ \ \ \ket{x_1} \otimes...\otimes D_{j, \sigma}\ket{x_j} \otimes ...  \otimes \ket{x_N},
\end{align}
where $x_j \in \{\textrm{vac}_{\textrm{H}},\uparrow_{\textrm{H}},\downarrow_{\textrm{H}},\downarrow\uparrow_{\textrm{H}}\}$ and $n_{j,\sigma} = d^{\dagger}_{j,\sigma} d_{,j\sigma}$. Hence, there is an additional factor of $(-1)^{n-1}$ that needs to be taken into account. With this, the equivalence Eq.\ \eqref{eq:hubbard_mapping} has been established.

Finally, we need to express the spin operators as well as the single-site projector $P_{j,\textrm{BCS}}$ in terms of the $d$ operators. As shown in the previous section, the spins are now simply the spins in the Hubbard model and hence we have
\begin{subequations}
\begin{align}
    S^z_{j} \sim \frac{1}{2}\pqty{d_{j, \uparrow}^{\dagger} d_{j, \uparrow} -  d_{j \downarrow}^{\dagger} d_{j \downarrow} } \equiv S^z_{j} ,\ 
    S^+_{j} \sim&\  d_{j, \uparrow}^{\dagger} d_{j, \downarrow} \equiv S^+_{j},\ 
    S^-_{j} \sim  d_{j, \downarrow}^{\dagger} d_{j, \uparrow} \equiv S^-_{j}, \\
     P_{j,\textrm{BCS}} \sim&\  \sum_{\sigma} d_{j, \sigma}^{\dagger} d_{j, \sigma} \equiv \sum_{\sigma} n_{j,\sigma}. 
\end{align}
\end{subequations}
With this we can write the effective model,
\begin{align}
    H_{tJ} = N \pqty{V - \frac{3}{4}K} + \sum_j \Bigg\{& - E_{\textrm{YSR}} \sum_{\sigma} n_{j,\sigma} + U n_{j,\uparrow} n_{j,\downarrow} +  \mathbf{S}_j \cdot J \cdot  \mathbf{S}_{j+1} \nonumber \\
    &-\tilde{t}\sum_{\sigma}\pqty{d^{\dagger}_{j,\sigma} d_{j+1,\sigma} + \textrm{h.c.} } - \tilde{\Delta} \pqty{ d^{\dagger}_{j,\uparrow}d^{\dagger}_{j+1,\downarrow} - d^{\dagger}_{j,\downarrow}d^{\dagger}_{j+1,\uparrow} + \textrm{h.c.} }  \Bigg\},
\end{align}
where $U \to \infty$. This is similar to a $t-J$ model (no double occupation, nearest-neighbor spin interaction), but explicitly includes a pairing term. Furthermore, the spin interaction is not necessarily small and may be ferromagnetic. We hence refer to this model as an extended $t-J$ model. Periodic boundary conditions in the original model translate to periodic boundary conditions in the extended $t-J$ model for $N$ even, and to antiperiodic boundary conditions for $N$ odd (for both hopping and pairing terms). This is due to the factor of $(-1)^1 (-1)^{N}$ arising from products of $\gamma_{N}$ and $\gamma_{1}$ and conjugates and corresponds to a $\pi$ flux threading the infinite-$U$ Hubbard ring for odd-$N$ chains. 

\subsection{Spin-orbit coupling in the effective $t-J$ model}\label{sec:supp_ext_t-J_soc}

We derive the terms in the extended $t-J$ model, when spin-orbit coupling is added to the original model in Eq.\ \eqref{eq:model}. We assume that the spin chain is aligned along the $x$-direction with the normal to the surface pointing along the $z$-direction. Then, the Rashba spin-orbit term is 
\begin{align}\label{eq:ham_soc}
	 H_{\textrm{SOC}} = -i t \alpha \sum_{j} \pqty{ c^{\dagger}_{j}\sigma_y c_{j+1} - \textrm{h.c.} }. 
\end{align}
Expressing this in terms of the local Bogoliubov operators given in Eq.\ \eqref{eq:bogoliubov_operators} and performing the replacement Eq.\ \eqref{eq:hubbard_mapping}, we obtain 
\begin{align}\label{eq:ham_ising_rashba_soc}
	 H_{\textrm{SOC},tJ} = \alpha   \sum_{j} \bqty{  -i \tilde{t} \pqty{ d^{\dagger}_{j}\sigma_y d_{j+1} - \textrm{h.c.} } + \tilde{\Delta} \pqty{ d^{\dagger}_{j}\sigma_0 d^{\dagger}_{j+1} + \textrm{h.c.} }  }. 
\end{align}
Thus, spin-orbit-coupling induces $p$-wave pairing in the extended $t-J$ model. Note that in the absence of spin order, this does not suffice to open a topological gap. To see this, consider the sum of the hybridization and spin-orbit terms. This may be written as 
\begin{align} \label{eq:ham_ising_rashba_soc2}
    H_{t,tJ} + H_{\textrm{SOC},tJ} = \sum_j\pqty{ - d^{\dagger}_{j} T_{\textrm{H}} d_{j+1} + d^{\dagger}_{j} \Delta_{\textrm{H}} d^{ \dagger}_{j+1}  +\textrm{h.c.} },
\end{align}
with $T = \tilde{t}(1 + i\alpha \sigma_y)$ and $\Delta_H = \tilde{\Delta}(1 + i\alpha \sigma_y)(i\sigma_y)$. The tunneling and pairing matrices are phase-locked and no gap opens in the absence of the spin interactions.

\subsection{Spectral function}\label{sec:supp_spectral_function}

Tunneling spectroscopy probes the local single-particle spectral function
\begin{align}
    A_{j,\sigma,a} (E) = \sum_{\lambda} \abs{\bra{\lambda} \psi^{\dagger}_{j,\sigma,a} \ket{\textrm{g.s.}} }^2 \delta (E - E_{\lambda} + E_0 ) , 
\end{align}
where we define $\psi_{j,\sigma,a} = ( c_{j,\sigma} , c^{\dagger}_{j,\bar{\sigma}})_a$ to treat tunneling in and out (or: tunneling of electron and hole) on the same footing ($a = 1 \equiv$ ``in", and $a = 2 \equiv$ ``out"). Expressing $\psi_{j,\sigma}$ in terms of local Bogoliubov operators and using the mapping to the extended $t-J$ model, we have
\begin{align}
    \psi_{j,\sigma} =
    \begin{pmatrix}
    u \\ \sigma v
    \end{pmatrix} \gamma_{j, \sigma}
    +
    \begin{pmatrix}
    -\sigma v \\ u
    \end{pmatrix} \gamma^{\dagger}_{j, \bar{\sigma}} 
    \sim  
    \frac{(-1)^{j}}{\sqrt{2}}\bqty{
    \begin{pmatrix}
    \sigma u \\  v
    \end{pmatrix} d^{\dagger}_{j, \bar{\sigma}}
    +
    \begin{pmatrix}
    v \\ - \sigma u
    \end{pmatrix} d_{j, \sigma} }.
\end{align}
Hence, up to a global sign, tunneling in of $\sigma$-electron and tunneling out of $\bar{\sigma}$-electron differ by $u \to \sigma v, v \to - \sigma u$. With this, the spectral function can be expanded as 
\begin{align}
    A_{j,\sigma} (E) = \frac{1}{2}\sum_{\lambda} 
    \Bigg\{&\ 
    \abs{\bra{\lambda} \gamma^{\dagger}_{j,\sigma} \ket{\textrm{g.s.}} }^2  
    \begin{pmatrix}
    u^2 \\  v^2
    \end{pmatrix} 
    + \abs{\bra{\lambda} \gamma_{j,\bar{\sigma}} \ket{\textrm{g.s.}} }^2  
    \begin{pmatrix}
    v^2 \\  u^2
    \end{pmatrix} 
    \\ &+2uv \Re\bqty{ \bra{\lambda} \gamma^{\dagger}_{j,\sigma} \ket{\textrm{g.s.}}
    \bra{\textrm{g.s.}} \gamma^{\dagger}_{j,\bar{\sigma}} \ket{\lambda} }\begin{pmatrix}
    - \sigma \\  \sigma
    \end{pmatrix} 
    \Bigg\} \delta (E - E_{\lambda} + E_0 ),
\end{align}
or in terms of the extended $t-J$ model
\begin{align}
    A_{j,\sigma} (E) = \frac{1}{2}\sum_{\lambda} 
    \Bigg\{&\ 
    \abs{\bra{\lambda} d^{\dagger}_{j,\sigma} \ket{\textrm{g.s.}} }^2  
    \begin{pmatrix}
    v^2 \\  u^2
    \end{pmatrix} 
    + \abs{\bra{\lambda} d_{j,\bar{\sigma}} \ket{\textrm{g.s.}} }^2  
    \begin{pmatrix}
    u^2 \\  v^2
    \end{pmatrix} 
    \\ &+2uv \Re\bqty{ \bra{\lambda} d^{\dagger}_{j,\sigma} \ket{\textrm{g.s.}}
    \bra{\textrm{g.s.}} d^{\dagger}_{j,\bar{\sigma}} \ket{\lambda} }\begin{pmatrix}
    \sigma \\  -\sigma
    \end{pmatrix} 
    \Bigg\} \delta (E - E_{\lambda} + E_0 ).
\end{align}
The states are now to be considered as eigenstates of the extended $t-J$ Hamiltonian $H_{\textrm{eff}}$. The first and second terms correspond to the electron and hole spectral functions of the extended $t-J$ model, the last corresponds to the spectral function associated with the anomalous Green function. In the main text, we specify to tunneling in of a spin-up electron. In the case of spin-ordered degenerate ground states (XXZ ferromagnets), we accordingly consider tunneling into the spin-down polarized ground state. To obtain the tunneling spectra in Fig.\ \ref{fig:heisenberg} (Fig. \ref{fig:ising_with_soc}), we calculate the matrix elements using the exact-diagonalization wave functions and approximate the $\delta$-function as a Gaussian peak with width $\kappa = 0.014 t$ ($\kappa = 0.0036t$). Finally, note that the spectra are normalized with respect to their individual maximal values $A_{\textrm{max}} = \max_{E,j} A_{j,\uparrow,\textrm{in}} (E)$. Thus, the magnitude should not be compared between different plots. 

\section{Mean-field theory of the metallic ferromagnet and topological superconductivity}
 
\begin{figure}
\centering
\includegraphics[width=0.98\textwidth]{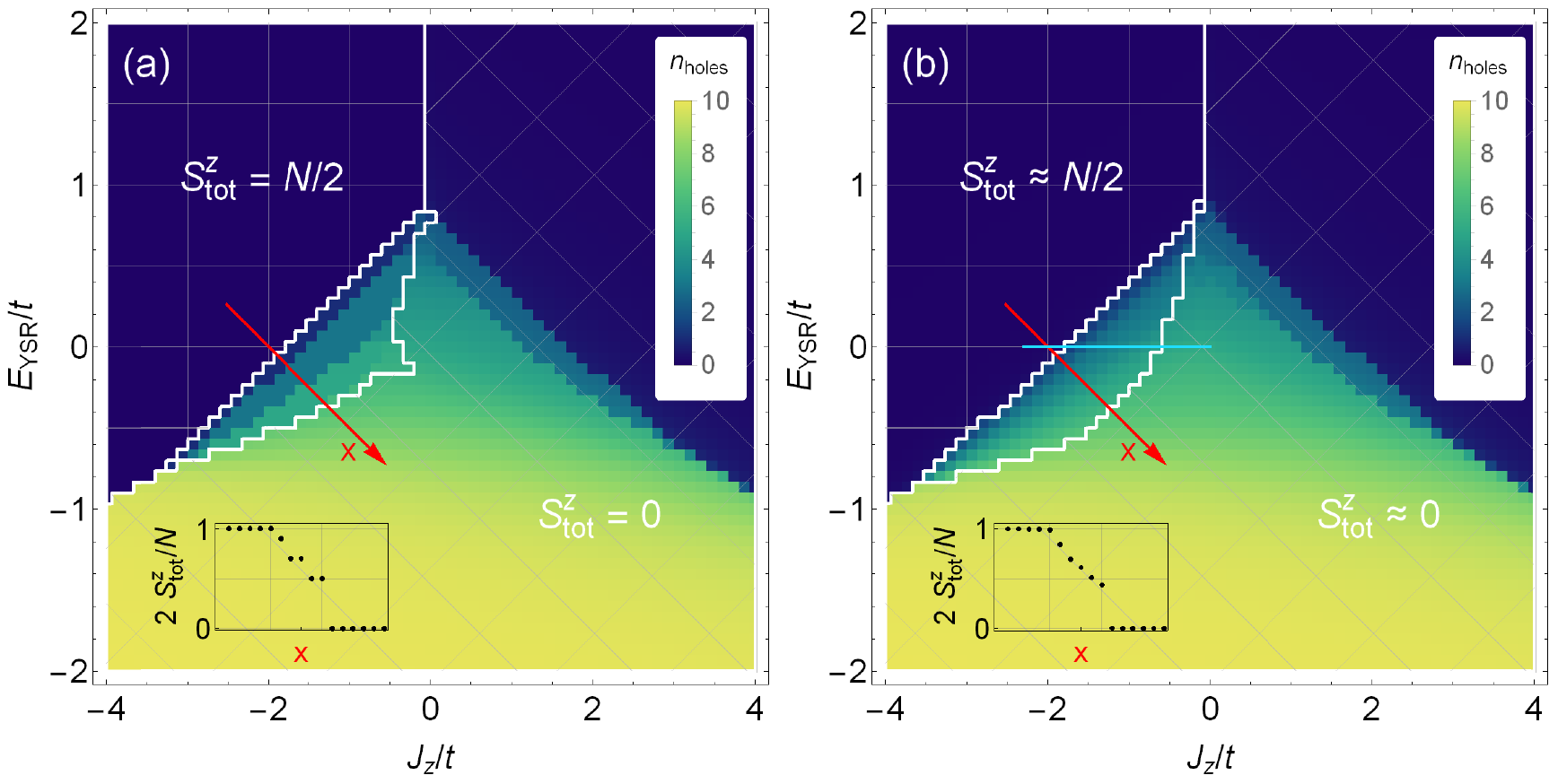}
     \caption{Exact-diagonalization results for $N=10$ chains with Ising RKKY coupling [$J = \textrm{diag}(0,0,J_z)$] and periodic boundary conditions, (a) without and (b) with spin-orbit coupling. In (a), the white lines delineate regions of maximum (minimum) $S^z_{\textrm{tot}}$. In (b), $S^z_{\textrm{tot}}$ is no longer a good quantum number due to spin-orbit coupling. Here, the white lines delineate regions for which $S^z_{\textrm{tot}}$ is within $0.2$ of its extremal values $0$ and $5$. For both (a) and (b), the rough features of the phase diagram are qualitatively similar to the phase diagram for isotropic RKKY interactions shown in the main text in Fig.\ \ref{fig:heisenberg}(b). On the ferromagnetic side $J_z < 0$, (a) exhibits the metallic ferromagnet (enclosed by the white lines). Across this region, the spin is reduced from its maximal value until the phase boundary to the singlet superconductor is reached (inset). The most striking difference in (b) is the breaking of particle number conservation in the ferromagnetic phases (spin-chain phase and metallic ferromagnet), which reflects the triplet contribution to the pairing. The light blue cut corresponds to the parameter range shown in Figs.\ \ref{fig:ising_with_soc} (a) and (b) of the main text. Parameters: $V = 2\Delta$, $B_z = -10^{-3}t$, (b) $\alpha = 0.25$.}
    \label{fig:appendix_ising}
\end{figure}

Topological superconductivity appears in the model of Eq.\ \eqref{eq:model} in the parameter region of the metallic ferromagnet, c.f. main text and Fig. \ref{fig:heisenberg} (b). When amending the model by spin-orbit coupling, a $p$-wave gap can open in this region. As shown in Sec.\ \ref{sec:supp_ext_t-J_soc}, spin-orbit coupling at the level of the model Eq. \eqref{eq:model} introduces a $p$-wave-pairing term into the extended $t-J$ model, which does not open a gap in the absence of some form of spin order.  According to the Mermin-Wagner theorem, there cannot be spin order for isotropic interactions \cite{Stoudenmire2011}. Hence, we explicitly break the spin rotation symmetries by considering XXZ coupling $J = \textrm{diag}(J_\perp,J_\perp,J_z)$. Moreover, the spin-orbit coupling, Eq. \eqref{eq:ham_soc}, breaks the SU(2) spin rotation symmetry down to a U(1) symmetry corresponding to rotations about the $y$-axis). In the numerics, we specified to the extreme case of an Ising interaction for simplicity, but the arguments below only require $J_z > J_\perp$. Figure \ref{fig:appendix_ising}(a) shows that the qualitative features of the phase diagram for Heisenberg interactions are robust against breaking the spin rotation symmetry. In particular, the ferromagnetic metal phase is maintained. With spin-orbit coupling, see Fig.\ \ref{fig:appendix_ising}(b), this phase becomes a topological superconductor as we will discuss in more detail below (see also Fig.\ \ref{fig:ising_with_soc} in the main text). 

\subsection{Metallic ferromagnet}

The metallic ferromagnet can be accurately described within a simple Hartree-Fock approach. We specify to ferromagnetic XXZ coupling (with $J_z\geq J_\perp$) and focus on states with maximal spin projection. As double occupation is forbidden, we use a spin-polarized Fermi sea as a variational Hartree-Fock ground state, 
\begin{equation}\label{eq:hartree_fock_state_fmm}
    \ket{\textrm{FS},\uparrow} = \prod_{\abs{k} < k_F} d_{k,\uparrow}^{\dagger} \ket{\textrm{vac}},
\end{equation}
in terms of $d_{j,\sigma} = \frac{1}{\sqrt{N}}\sum_k d_{k,\sigma}e^{ikj}$ with $k=\frac{2\pi n}{N} \in [-\pi,\pi]$ and $n\in\mathbb{Z}$ (periodic boundary conditions). The Fermi momentum $k_F$ can be viewed as a variational parameter and the number of holes (screened sites) is given by  
\begin{equation}
    n_\mathrm{holes} = N - \sum_{\abs{k} < k_F} \bra{\textrm{FS},\uparrow}  d^\dagger_{k, \uparrow} d_{k, \uparrow} \ket{\textrm{FS},\uparrow} = N\left( 1- \frac{k_F}{\pi}\right).
\end{equation}
Here, the last term on the right hand side takes the thermodynamic limit $N\to\infty$. As the singlet pairing $\tilde\Delta$  does not contribute, the Hamiltonian in Eq.\ (2) implies that this state has energy
\begin{equation}
   E_{\mathrm{FS}}(k_F) = \bra{\textrm{FS},\uparrow} H \ket{\textrm{FS},\uparrow} = \sum_{\abs{k}<k_F} (- E_{\textrm{YSR}} - 2{\tilde t}\cos k ) + \bra{\textrm{FS},\uparrow} \sum_{j} \mathbf{S}_j \cdot J \cdot \mathbf{S}_{j+1}  \ket{\textrm{FS},\uparrow} .
\end{equation}
Only the longitudinal exchange coupling $J_z$ contributes to the RKKY interaction and we obtain (for $N\to\infty$)
\begin{equation}
    \bra{\textrm{FS},\uparrow} \sum_{j} \mathbf{S}_j \cdot J \cdot \mathbf{S}_{j+1}  \ket{\textrm{FS},\uparrow} = \frac{N J_z}{4\pi^2} (k_F^2 - \sin^2 k_F).
\end{equation}
Altogether, we find the energy  
\begin{align}\label{eq:hartree_fock_energy}
    \frac{E_{\mathrm{FS}}(k_F)}{N} 
    =&\ - \frac{k_F}{\pi} E_{\textrm{YSR}} - \frac{2\tilde t}{\pi} \sin k_F + \frac{J_z}{4\pi^2} \pqty{k_F^2 - \sin^2 k_F} \nonumber \\
    =&\  \int_{-k_F}^{k_F} \frac{\textrm{d}k}{2\pi}\  \hat{\xi}(k) - \frac{J_z}{4\pi^2} \pqty{k_F^2 - \sin^2 k_F} ,
\end{align}
where we defined the mean-field dispersion 
\begin{equation}
\hat{\xi}(k) = -E_{\textrm{YSR}} -2 \tilde{t}\cos k + J_z \pqty{ \frac{k_F}{2\pi}  - \frac{\sin k_F}{2\pi} \cos k }.
\end{equation}
Minimizing the energy with respect to $k_F$, we find an implicit expression for $k_F$ and hence the number of holes as a function of the YSR energy $E_\mathrm{YSR}$ and RKKY interaction $J_z$, 
\begin{equation}\label{eq:hartree_fock_energy_minimize}
    0 = \hat{\xi}(k_F).
\end{equation}
In particular, this expression with $k_F=\pi$ implies that the transition between the Heisenberg spin chain ($n_\mathrm{holes}=0$) and the metallic ferromagnet occurs for
\begin{equation}
   E_\mathrm{YSR} = 2\tilde t + \frac{J_z}{2}.
   \label{upper}
\end{equation}
This expression can also be understood by noting that the first hole enters at the top of the band at energy $2\tilde t$ and breaks two ferromagnetic bonds. Similarly, we find a transition to a fully screened chain ($k_F=0$) at
\begin{equation}
    E_\mathrm{YSR} = - 2\tilde t 
\label{lower}
\end{equation}
for negligible pairing $\tilde \Delta$. Nonzero pairing preempts the transition from the metallic ferromagnet into the fully screened chain by the formation of a singlet superconducting phase at a finite density of $d$ fermions. 
This is not captured by the ansatz Eq.\ \eqref{eq:hartree_fock_state_fmm}. 

\subsection{Topological superconductivity}

\begin{figure*}
\centering
\includegraphics[width=0.98\textwidth]{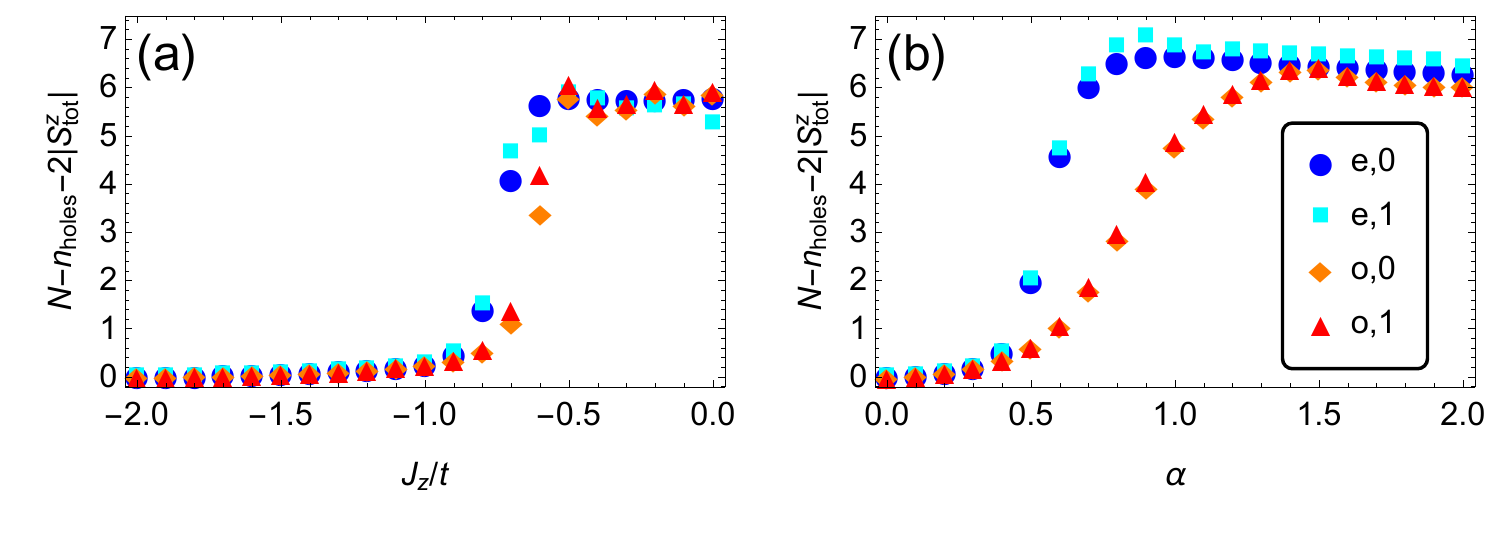}
    \caption{
    Exact-diagonalization data for the deviation from full spin polarization $N - n_{\textrm{holes}} - 2 \abs{S^z_{\textrm{tot}}}$ in a spin-orbit-coupled chain with $N=12$ sites, with Ising RKKY interactions and open boundary conditions. 
    Blue and cyan markers correspond to two lowest-energy even-parity states (circle, square; labeled by e,0/1), orange and red markers correspond to two lowest-energy odd-parity states (diamond, triangle; labeled by o,0/1). 
    These four states make up the ground-state manifold in Fig.\ \ref{fig:ising_with_soc}(a). 
    The system is approximately spin polarized (deviation from full spin polarization $\simeq 0$) for sufficiently large $J_z$ and sufficiently small $\alpha$. 
    The latter requirement is due to the tendency of spin-orbit coupling to suppress the spin polarization of the metallic ferromagnet.  
    This delineates the parameter range, for which the spin polarized variational theory developed in this section is adequate. Parameters: $E_{\textrm{YSR}} = 0$, (a) $\alpha = 0.25$, (b) $J_z = -1.25t$. 
    As in Fig.\ \ref{fig:ising_with_soc}, a small magnetic field $B_z = 10^{-3}t$ along the $z$-direction was added for numerical reasons.  
    }
    \label{fig:spin_polarizedness}
\end{figure*}

The metallic ferromagnet serves as a parent state for topological superconductivity, which can develop in the presence of spin-orbit coupling. We choose Rashba spin-orbit coupling as defined in Eq.\ \eqref{eq:ham_soc}. To obtain an approximate theory for the topological superconducting phase, we start from the spin-polarized metallic state introduced in the previous section (with $k_F$ determined by $\hat{\xi}(k_F)=0$) and add $p$-wave pairing $\Delta_{\textrm{p}}$. We will discuss the precise form of $\Delta_{\textrm{p}}$ below. This assumes that the ground state remains approximately spin polarized in the presence of weak spin-orbit coupling. This is a good approximation for $J_z \lesssim - 1$ and $\alpha \lesssim 0.4$ as can be inferred from Fig.\ \ref{fig:spin_polarizedness}.  The ground-state energy density is
\begin{equation}
    \frac{E_{\textrm{TSC}}}{N} =\frac{1}{2} \int_{-\pi}^{\pi} \frac{\textrm{d}k}{2\pi} \bqty{ \hat{\xi}(k;k_F)   - \sqrt{
    \hat{\xi}^2(k;k_F) 
    +\Delta^2_{\textrm{p}}(k;k_F) } } -  \frac{J_z}{4\pi^2} \pqty{k_F^2 - \sin^2 k_F}.
\end{equation}
and the total spin along the $z$-direction is 
\begin{align}\label{eq:pwave_mf_magnetization}
    S^z_{\textrm{tot}} = \frac{N}{4}\bqty{1 -\int_{-\pi}^{\pi} \frac{\textrm{d}k}{2\pi} \frac{\hat{\xi}(k;k_F) }{\sqrt{  \hat{\xi}^2(k;k_F) + \Delta^2_{\textrm{p}}(k;k_F) }} }.
\end{align}
The excitation gap is given by
\begin{equation}\label{eq:pwave_mf_gap}
    E_{\textrm{gap}} = \min_k \sqrt{
    \hat{\xi}^2(k;k_F) 
    + \Delta^2_{\textrm{p}}(k;k_F) }. 
\end{equation}
It remains to discuss the $p$-wave pairing. There are two contributions to $\Delta_{\textrm{p}}(k)$ in the presence of Rashba spin-orbit coupling ($\propto \alpha \sigma_y$): a direct contribution derived in Sec.\ \ref{sec:supp_ext_t-J_soc} and a contribution mediated by virtual spin flips and singlet pairing. The latter is equivalent to the mechanism by which spin-orbit-coupled nanowires proximity couple to an $s$-wave superconductor and acquire a $p$-wave gap (in the strong-field limit). To derive this contribution, we assume a large mean field along the positive $z$-direction, 
\begin{equation}
    \frac{J_z S^z_{\textrm{tot}} }{N} \simeq \frac{J_z k_F}{2\pi}.
\end{equation}
Treating the pairing terms in perturbation theory and expanding for small $\alpha\tilde{t}/J_z k_F$, we obtain the the effective $p$-wave pairing 
\begin{equation}\label{eq:pwave_pairing}
    \Delta_{\textrm{p}}(k;k_F) = 4 i \tilde{\Delta} \alpha \sin k \bqty{1 + \frac{2\pi\tilde{t}}{J_z k_F}\cos k}.  
\end{equation}
The results of this section are illustrated in Fig.\ \ref{fig:ising_with_soc} along with the exact diagonalization data. Panel (a) includes the $p$-wave gap as determined from Eqs.\ \eqref{eq:pwave_mf_gap} and \eqref{eq:pwave_pairing} (black, dot-dashed), while panel (b) includes $\abs{S^z_{\textrm{tot}}}$ and $n_{\textrm{holes}}$ ($ = \abs{S^z_{\textrm{tot}}}/2$ within the spin-polarized ansatz) determined from Eqs.\ \eqref{eq:pwave_mf_magnetization}. $k_F$ is determined by minimizing the normal state energy, Eq.\ \eqref{eq:hartree_fock_energy_minimize}.

\end{document}